\documentclass[%
 reprint,
 superscriptaddress,
 amsmath,amssymb,
 aps,
]{revtex4-2}

\usepackage{graphicx}
\usepackage{dcolumn}
\usepackage{bm}
\usepackage{braket}
\usepackage{xcolor}
\usepackage{amsmath}
\usepackage{upgreek}
\usepackage{multirow} 
\usepackage{textalpha}
\usepackage{xr}  

\begin{document}

\preprint{APS/123-QED}

\title{Single-Scan Characterization of $^{14}$N Nuclei via $^1$H-Detected Rotating-Frame Relaxometry}%

\email[Contact author: ]{alexej.jerschow@nyu.edu}

\author{Florin Teleanu}
\affiliation{Department of Chemistry, New York University, New York, NY 10003, United States}
\affiliation{ELI-NP, “Horia Hulubei” National Institute for Physics and Nuclear Engineering, 30 Reactorului Street, Bucharest-Magurele, 077125, Ilfov, Romania}

\author{Huijing Zou}
\affiliation{Department of Chemistry, New York University, New York, NY 10003, United States}

\author{David E. Korenchan}
\affiliation{A. A. Martinos Center for Biomedical Imaging, Massachusetts General Hospital, Charlestown, MA,  United States}
\affiliation{Harvard Medical School, Boston, MA, United States}
\author{Alexej Jerschow*\textsuperscript{,}}
\affiliation{Department of Chemistry, New York University, New York, NY 10003, United States}

\date{\today}

\begin{abstract}
$^{14}$N NMR is notoriously difficult to perform in liquids due to the  very fast spin relaxation and the large quadrupolar couplings, which render many signals invisible.   We show here how $^{14}$N nuclei of biomolecular constituents can be probed indirectly by reintroducing the scalar relaxation of the second kind contribution to the polarization lifetimes of J-coupled protons in double resonance spin-locking experiments. The enhanced $^1$H relaxation rates in the rotating-frame allow for direct evaluation of nitrogen chemical shift and polarization lifetimes, from which one- and even two-bond $^1$H-$^{14}$N scalar couplings as well as $^{14}$N quadrupolar interactions can be determined. We demonstrate the versatility of this method by characterizing $^1$H-$^{14}$N spin pairs in several molecules of biological importance, showing proton relaxation enhancements beyond one order of magnitude. We further observe a pronounced effect from  intermolecular hydrogen bonding. Our approach can be readily integrated into existing biomolecular NMR methodologies, as demonstrated here for $^1$H-detected relaxation-editing experiments with water suppression. This method provides access to nitrogen's picosecond-modulated quadrupolar interaction via single-scan proton detection in systems that would otherwise yield almost no detectable direct $^{14}$N signal even after averaging over thousands of transients.
\end{abstract}

\maketitle

\section*{Introduction}

Biomolecular NMR has become a central tool for structural characterization and for probing dynamic processes such as ligand binding\cite{buchanan_pathogen-sugar_2022,sprangers_quantitative_2007}, protein folding\cite{kay_backbone_1989,tollinger_slow_2001}, and nucleic acid stability\cite{zhang_visualizing_2007,nikolova_transient_2011}. Accurate investigation of biomolecules often requires expensive and time-consuming isotopic enrichment to replace naturally abundant NMR-inactive nuclei, such as $^{12}$C, or rapidly relaxing nuclei, such as $^{14}$N. Significant advances have been made in full or selective isotopic labeling with $^{13}$C and $^{15}$N nuclei\cite{hong_selective_1999,sugiki_amino_2018}, which allowed for better peak assignment via multidimensional correlation mapping and extended lifetime of nuclear spin transitions\cite{pervushin_attenuated_1997}. Furthermore, high-field relaxometry techniques enabled by isotopic enrichment of proteins/nucleic acids have granted access to the dynamical features of biomolecules on timescales ranging from fast ps-ns motion probed by $T_1$, $T_2$ relaxation measurements, to ns-$\upmu$s motion in $T_{1\rho}$ rotating-frame relaxation experiments and slow ms chemical exchange dynamics derived from relaxation dispersion techniques. 

Controlling relaxation pathways offers great opportunities for magnetic resonance spectral editing and polarization storage, yet only a few cases are known in which spin interactions can be tuned at will through tailored radio-frequency pulses or spin state preparation. As examples, we note, in particular, the acceleration of the polarization recovery in SOFAST experiments by selective excitation of a specific proton spectral region\cite{schanda_very_2005,farjon_longitudinal-relaxation-enhanced_2009}, the enhancement of spectral resolution by exploiting the constructive/destructive interference of dipolar and chemical shift anisotropy interactions in TROSY experiments\cite{pervushin_attenuated_1997,pervushin_transverse_1998}, and rendering the intra-pair homonuclear dipole-dipole interactions ineffective by the use of 
long-lived nuclear spin states\cite{carravetta_beyond_2004,pileio_long-lived_2020}. 

Here, we show how to control and exploit the scalar relaxation of the second kind (SR2K) contribution\cite{werbelow_nuclear_1997} impacting $^1$H polarization lifetimes in double resonance spin-locking experiments as an indirect measurement modality to characterize J-coupled $^{14}$N nuclei. Compared to other $^1$H-detected methods that only probe proton-nitrogen connectivity via $^{1}$H-$^{15}$N scalar couplings at natural abundance (0.38\%) such as HMBC experiments\cite{bax_correlation_1983,bodenhausen_correlation_1977}, our approach exploits the significant enhancement of proton relaxation rates in the rotating-frame which is maximized when both $^1$H and $^{14}$N nuclei are irradiated on resonance under the Hartmann-Hahn condition\cite{hartmann_nuclear_1962,rao_relaxation_1965,skrynnikov_efficient_1998}. The great advantage of this method is the enhanced sensitivity via proton detection (within one transient) of otherwise invisible peaks in the $^{14}$N NMR spectrum (averaged over several thousands transients due to very fast spin decoherence). This approach also enables the indirect characterization of the $^{14}$N quadrupolar interaction which can act as a reporter of  molecular dynamics. Similar strategies exploiting the SR2K contribution to characterize interactions involving fast relaxing quadrupolar nuclei are very limited in liquid-state NMR with only a few studies investigating $^{13}$C-$^{79}$Br spin systems\cite{levy_carbon-13_1972,gryff-keller_scalar_2014,kubica_scalar_2014,bernatowicz_scalar_2014}. As the two nuclei have very similar gyromagnetic ratios ($\gamma_{13\text{C}}/\gamma_{79\text{Br}}\approx1.0004$), the SR2K effect is readily observed in laboratory frame measurements of classical relaxation times $T_1$ and $T_2$, which cannot be controlled by radio-frequency pulses in this situation. The first theoretical derivation and experimental observation of this effect in the rotating-frame was done by Skrynnikov et al.\cite{skrynnikov_efficient_1998} showing significant relaxation enhancement for amine protons via double resonance irradiation. To our knowledge, only one study focused on deriving structural and dynamical features via the SR2K contribution in the rotating frame involving double-resonance spin-lock irradiation of $^{13}$C-$^{14}$N spin systems in biomolecules, leading to pulse-modulated relaxation contributions\cite{gryff-keller_scalar_2012}. Alternatively, there have been attempts to exploit the SR2K effect to modulate the water's $^1$H relaxation by irradiating $^{14}$N or $^{17}$O nuclei bearing exchangeable protons\cite{martinho_harnessing_2022} leading to promising new sources of relaxation contrast in magnetic resonance imaging. 

 \begin{figure*}[t]
  \centering
  \includegraphics[page=1,trim=00pt 240pt 00pt 00pt, clip, width=\textwidth]{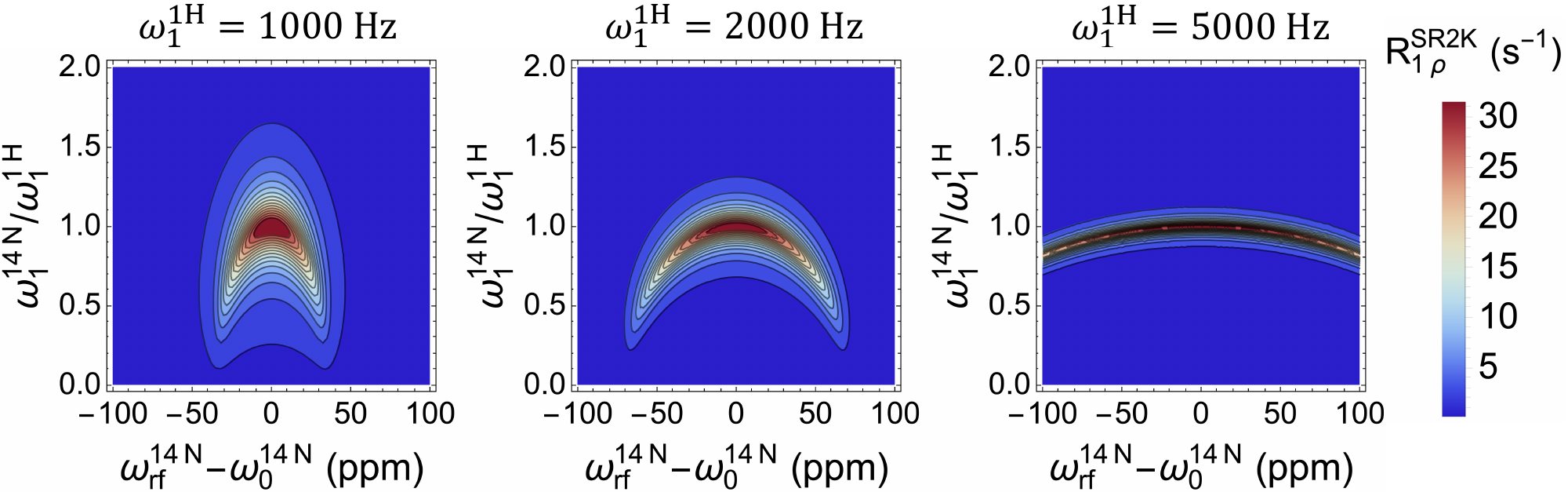}\caption{Numerical simulations of the impact of the $^{14}$N carrier offset ($\omega_{rf}^{14\text{N}}-\omega_{0}^{14\text{N}}$) and spin-lock power ratio ($\omega_{1}^{14\text{N}}/\omega_{1}^{1\text{H}})$ on the scalar relaxation of the second kind (SR2K) contribution to the $^1$H relaxation rate in the rotating frame ($R_{1\rho}^{\text{SR2K}})$ at different proton irradiation frequencies ($\omega_{1}^{1\text{H}}$). Plots are generated using equation 26 from Ref \cite{skrynnikov_efficient_1998} assuming on-resonance irradiation at the proton Larmor frequency ($\omega_{\text{rf}}^{1\text{H}}=\omega_{0}^{1\text{H}}$). The fixed numerical values used are $J_{HN}=65$\ Hz,  $T_\text{1}^{14\text{N}}=1$ms, $B_0=9.4$T.  }
  \label{theory}
\end{figure*}

The reintroduction of the SR2K contribution to proton relaxation in the rotating-frame $R_{1 \rho}$ due to a scalar coupled, fast relaxing $^{14}$N nucleus was first described by Skrynnikov\cite{skrynnikov_efficient_1998} who provided the following analytical equation:
\begin{widetext}
\begin{align}
R_{1\rho}^{SR2K}=\frac{1}{12}(2\pi J_{HN})^2 S(S+1)\left( J(T_1^{14N},\omega_{1}^{1H}-\omega_{1}^{14N})+J(T_1^{14N},\omega_{1}^{1H}+\omega_{1}^{14N})\right),
\label{Skrynikov}
\end{align}
\end{widetext}

where $J_{HN}$ is the scalar coupling constant, $S$ is the nuclear spin value of the fast relaxing nucleus ($S_{14N}=1$), $T_1^{14N}$ is the longitudinal polarization lifetime of $^{14}$N due to the quadrupolar interaction, $\omega_{1}^{1H/14N}$ are the nutation frequencies of $^1$H or $^{14}$N induced by resonant RF irradiation of both nuclei and $J(\tau,\omega)$ is the spectral density characterizing the modulation of the $^1$H-$^{14}$N scalar coupling interaction with a correlation time $\tau$ at a frequency $\omega$. Under the isotropic tumbling assumption of the Bloembergen-Purcell-Pound model\cite{bloembergen_relaxation_1948}, the spectral density takes the standard Lorentzian-shaped functional expression, leading to a maximum SR2K contribution for $\omega_1^{1H} = \omega_1^{14N}$. Thus, resonant irradiation at the Hartmann-Hahn condition\cite{hartmann_nuclear_1962} opens a new channel for proton relaxation\cite{skrynnikov_efficient_1998}. Figure \ref{theory} shows the predicted SR2K contribution to proton relaxation rate in the rotating frame $R_{1\rho}^{\text{SR2K}}$ as a function of nitrogen carrier offset  and mismatch of the spin-lock powers on the $^1$H and $^{14}$N channels. A double-well profile is apparent for off-resonance irradiation of the targeted nitrogen nucleus at $^{14}$N irradiation powers smaller than the one applied on the $^1$H channel. By increasing the irradiation power on the nitrogen channel, the maximum contribution is achieved for on-resonance irradiation of the $^{14}$N nucleus with carrier offset effects becoming less significant with increasing  spin-lock power.

At a first glace, Eq. \ref{Skrynikov} might suggest that a larger $T_{\text{1}}^{14\text{N}}$ should lead to a larger SR2K contribution. However, as pointed out by Skrynnikov\cite{skrynnikov_efficient_1998}, this equation is only applicable when
\begin{equation}
    2\pi J_{HN} T_1^{14N} < 1 < \omega_{1}^{14N}T_{1}^{14N}.
    \label{Skrynikov_cond}
\end{equation}
 Consequently, $^{14}$N polarization lifetimes should be short compared to the inverse of the scalar coupling constant, while the irradiation power (expressed in Hz) needs to be larger than the characteristic relaxation rates of the quadrupolar nucleus. Below the lower bound condition, the quadrupolar relaxation becomes too fast for the effect to be communicated via the J-coupling, and the higher bound describes the limit at which the quadrupolar relaxation is too fast for rf-power intervention (self-decoupling). Within these two limits, one can use Eq. \ref{Skrynikov} to fit the SR2K contribution to $^1$H relaxation rates in the rotating-frame measured in double resonance spin-locking experiments where the $^{14}$N irradiation power is steadily increased up to and beyond the Hartmann-Hahn condition.    The resulting relaxation dispersion profile will display a Lorentzian shape with full-width at half-height (FWHH) being inversely proportional to $T_{\text{1}}^{14\text{N}}$, while the maximum amplitude is proportional to $J{_{HN}^{2}}\cdot T_{\text{1}}^{14\text{N}}$. Consequently, by observing $^1$H $R_{1\rho}$ relaxation enhancement  as a function of $^{14}$N carrier frequency and $^{14}$N irradiation power levels, one can derive several spectral quantities characterizing the quadrupolar nucleus such as the chemical shift, the scalar coupling and the longitudinal polarization lifetime (and by extension the quadrupolar coupling and correlation time) even in cases where no $^{14}$N peak can be observed with standard NMR techniques due to very fast decoherence (see below). We validate this approach across 10 molecular systems, show its sensitivity to temperature changes and its possible utility to studying intermolecular hydrogen bond formation in systems where the proton is coordinated by two fast-relaxing nuclei. Lastly, we demonstrate the versatility and utility of this method by developing a new spectral editing technique that incorporates the dual channel spin-lock module into a $^1$H-detected water suppression experiment. 

  \begin{figure*}[t]
  \centering
  \includegraphics[page=2,trim=00pt 0pt 00pt 00pt, clip, width=\textwidth]{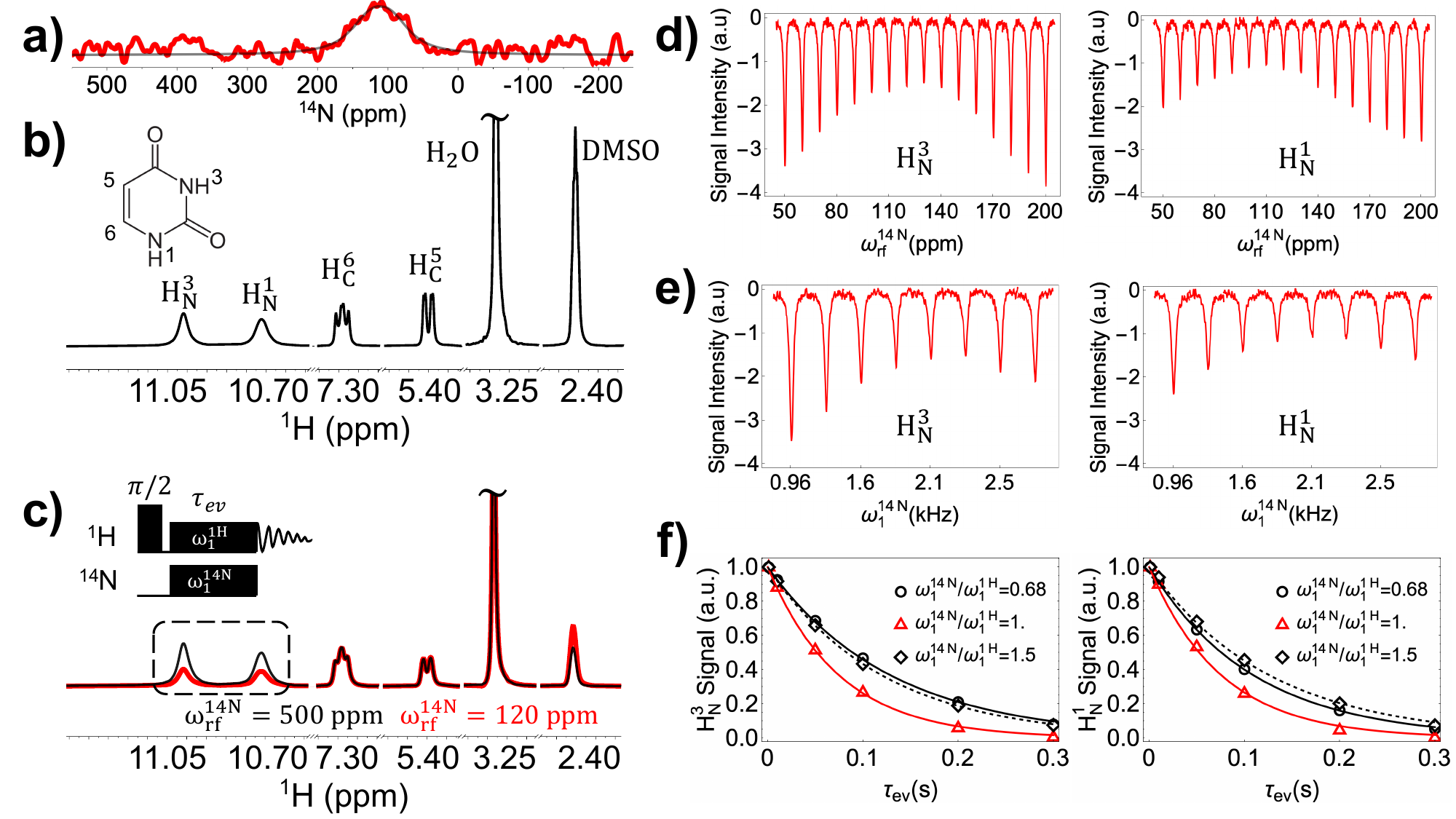}\caption{Indirect detection of $^{14}$N spins scalar coupled to $^1$H spins via $^1$H NMR relaxometry in the rotating-frame at 9.4 T for non-labeled uracil dissolved in deuterated DMSO. a) $^{14}$N spectrum measured using the aring sequence\cite{wang_triple-pulse_2021} averaged over 8192 transients (red) showing a very broad peak at around 115 ppm due to fast decoherence; Fitted spectrum (black) shows a FWHH of 2200 Hz (a line-broadening factor of 200 Hz was used) b) Standard pulse-acquire $^{1}$H spectrum of uracil (with peak assignment) showing traces of water and protonated DMSO. c) Spectra measured using the pulse sequence shown as inset with proton carrier set at $\omega_{\text{rf}}^{1\text{H}}=10.8$ ppm and matched nutation frequencies $\omega_1^{1\text{H}}=\omega_{1}^{14\text{N}}=2$ kHz for a constant irradiation time $\tau_{\text{ev}}=0.1$ s with on-resonance (red) and off-resonance (black) $^{14}$N irradiation (one transient); d) $^{14}$N carrier frequency scan and e) $^{14}$N nutation frequency scan optimized by monitoring H$_{\text{N}}^{3}$ and H$_{\text{N}}^{1}$ peak decrease  with the same values for the proton carrier and nutation frequencies and total irradiation time; f) $^1$H signal decay under resonant irradiation and different matching conditions of the nutation frequencies applied on the two heteronuclei. Experiments shown in (d-f) used the pulse sequence shown in (c).}
  \label{uracil}
\end{figure*}

\section*{Results and Discussion}
\label{results_discussion}
Although J-coupled $^1$H–$^{14}$N spin pairs are commonly found in biomolecules — for instance in the pyrimidine and purine heterocycles of nucleic acids or in the amine groups of amino acids — rapid proton exchange with the solvent renders these sites susceptible to line broadening. In aqueous solution, the corresponding proton signals are typically weak or undetectable owing to chemical exchange with water, which undermines the scalar coupling-driven relaxation enhancement. We therefore selected dimethyl sulfoxide (DMSO) and acetone as alternative polar solvents for our initial investigations. 

Figure \ref{uracil} summarizes the general workflow for reintroducing the SR2K contribution to the relaxation rates of amidic protons in uracil dissolved in deuterated DMSO. The $^{14}$N NMR spectrum shows a very broad peak around 115 ppm with poor signal-to-noise ration (SNR) despite acquiring more than 8000 scans (Fig. \ref{uracil}a) due to very fast spin relaxation which makes nitrogen scalar coupling and polarization lifetime measurements unreliable. The $^1$H spectrum shows clear detection and assignment of the uracil peaks (Fig. \ref{uracil}b) with amidic protons around 11 ppm. Using the pulse sequence shown in Fig \ref{uracil}c and setting the $^1$H carrier frequency of the spin-lock pulse to the center frequency between the  two amidic proton peaks, the SR2K contribution in the rotating-frame can be introduced during irradiation by applying a resonant power-matched spin-lock on the $^{14}$N channel. This manifests as a decrease in signal intensity of the irradiated NH protons after the spin-lock period $\tau_{ev}$ on both channels compared to the case of large $^{14}$N carrier offsets (Fig. \ref{uracil}c) where the SR2K contribution is inactive. The carrier frequency of the nitrogen spin-lock pulse can be estimated either from the observed peak in the $^{14}$N spectrum or, more reliably, by monitoring the change in the amidic proton signal while scanning the carrier frequency of the $^{14}$N irradiation pulse (Fig. \ref{uracil}d). The two proton peaks show a maximum drop for the $^{14}$N carrier set to around 130 ppm and 110 ppm, respectively. Notably, this small chemical shift separation cannot be resolved from direct $^{14}$N measurements (Fig. \ref{uracil}a). Then, the power matching condition can be monitored by setting a fixed $^1$H irradiation power ($\omega_{1}^{1\text{H}}=2$kHz) and performing experiments with increasing $^{14}$N spin-lock power. Results in Fig. \ref{uracil}e shows slight deviations from the expected minimum proton signal at $\omega_{1}^{14\text{N}}=2$kHz which might be due to pulse imperfections or possible uracil-DMSO association through hydrogen bonding that might lead to additional chemical exchange phenomena. Finally, we monitor the $^1$H signal decay during resonant irradiation on both channels as a function of irradiation time $\tau_{ev}$, highlighting significant relaxation enhancement for the Hartmann-Hahn condition $\omega_{1}^{1\text{H}}=\omega_{1}^{14\text{N}}$ (Fig. \ref{uracil}f) as predicted by theory.

\begin{figure*}[t]
  \centering
  \includegraphics[page=3,trim=00pt 160pt 00pt 35pt, clip, width=\textwidth]{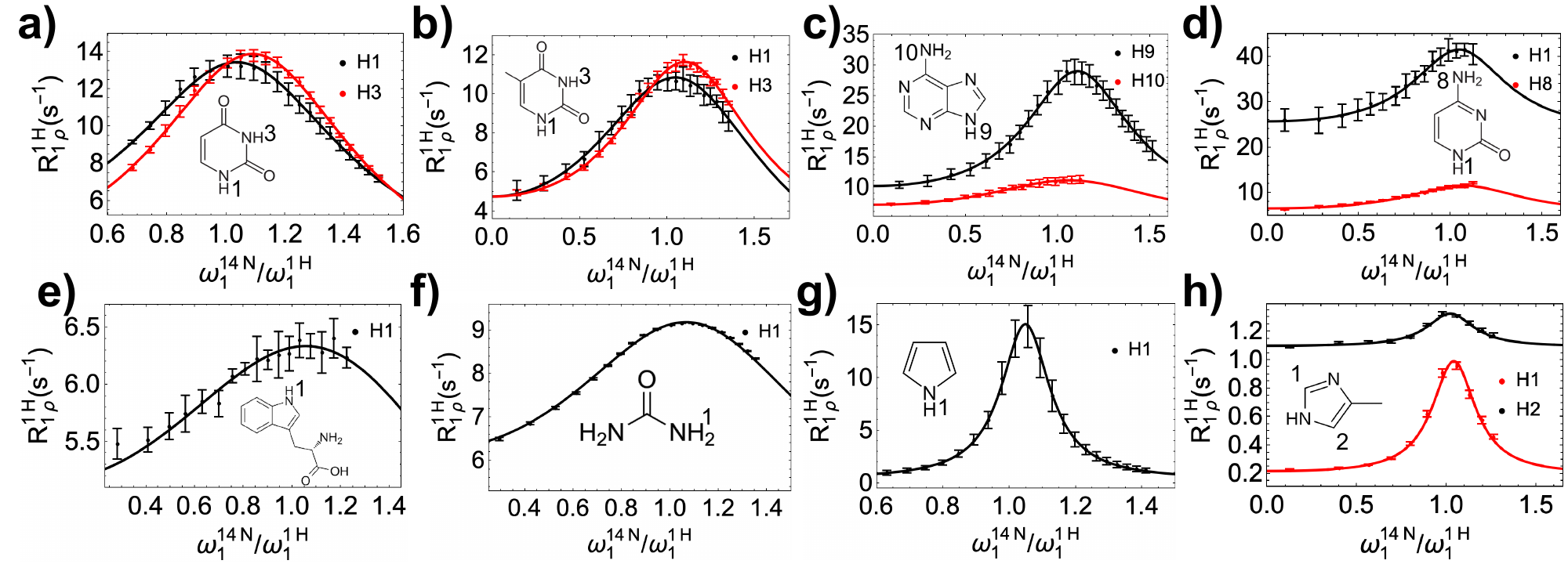}\caption{Rotating-frame $^1$H relaxation rates in double resonance spin-lock experiments with variable $\omega_{1}^{14\text{N}}/\omega_{1}^{1\text{H}}$ matching condition performed on uracil (a), thymine (b), adenine (c), cytosine (d), tryptophan (e), urea (f), pyrrole (g) and 4-methyl imidazole (h). For each experiment, the carrier frequencies on the two $^1$H and $^{14}$N channels were set on the targeted resonances as detailed in Fig. \ref{uracil}, while the irradiation power on the $^1$H channel was fixed (see Table II) and the $^{14}$N irradiation power was varied. The first six samples are prepared in DMSO-d6, while the last two are prepared in acetone-d6. Points represent experimentally-derived $^1$H relaxation rates (error bars shown) and lines represent fits based on Eq. \ref{Skrynikov} (see fitted parameters in Table \ref{table:fitted_data}).}
  \label{all_molec}
\end{figure*}

Having identified the targeted amino/amido $^1$H peaks and the optimal $^{14}$N carrier frequency of the irradiation pulse for all molecular systems by the same workflow (Table II), we conducted rotating-frame $^1$H relaxation measurements using the same pulse sequence (Fig. \ref{uracil}c) with fixed $^{1}$H and variable $^{14}$N irradiation powers. The relaxation dispersion profiles have been measured for multiple protons of the same molecule (Fig. \ref{all_molec}) and fitted with Eq. \ref{Skrynikov} from which $^1$H-$^{14}$N scalar coupling constants and $^{14}$N polarization lifetimes have been derived (Table \ref{table:fitted_data}). The four investigated nucleobases, uracil, thymine, adenine and cytosine, show scalar relaxation enhancements for the amido protons between 2- and 3-fold, with amino protons displaying smaller enhancements (Fig. \ref{all_molec}a-d). The fitted parameters are consistent across these molecules with one-bond $^1$H-$^{14}$N scalar couplings between 60-80 Hz and $^{14}$N relaxation times below 1ms. 

Then, we investigated the SR2K effect of indole's nitrogen nucleus in tryptophan and amidic nitrogen in urea showing modest scalar relaxation contributions (Fig. \ref{all_molec}e-f). In the case of pyrrole dissolved in acetone-d6 (Fig. \ref{all_molec}g), the maximum relaxation enhancement was 20-fold and the dispersion profile was significantly sharper compared to other systems suggesting a longer $^{14}$N polarization lifetime confirmed by the narrow linewidth of the $^{14}$N peak (Fig. 7) which allowed us to measure directly the longitudinal relation time as $T_{1}^{14\text{N}} \approx 5$ ms. However, given the readily-observable $J_{HN}\approx 65$ Hz in the proton spectrum, the dispersion profile cannot be accurately fitted with Eq. \ref{Skrynikov} as the spin system violates the condition of Eq. \ref{Skrynikov_cond} with $2\pi J_{HN} T_{1}^{14\text{N}}\approx 2$. Remarkably, the SR2K contribution was detectable for protons located two bonds away from the nitrogen atoms in 4-methylimidazole (Fig. \ref{all_molec}h), with the largest relaxation enhancements (5-fold) observed for the $^1$H nucleus coupled to both $^{14}$N nuclei simultaneously. This finding points to the possibility of extending the SR2K effect to molecular systems beyond those bearing N–H motifs — which are susceptible to proton exchange in protic solvents — and further highlights the cumulative nature of the SR2K contribution when multiple neighboring $^{14}$N nuclei are present.

\begin{table*}[t]
\centering
\caption{\small{Indirect $^{14}$N characterization via SR2K contribution to the $^1$H relaxation rates: nitrogen chemical shifts ($\omega_{rf}^{14\text{N}}$) are measured by scanning the $^{14}$N carrier frequency and monitoring the largest drop in the corresponding $^1$H signal; $^1$H-$^{14}$N scalar coupling constants ($J_{HN}$) and $^{14}$N longitudinal polarization lifetimes ($T_1^{\text{14N}}$) are derived by fitting the experimental rotating-frame $^1$H relaxation rate dispersion profiles (Fig. \ref{all_molec}) as a function of the $\omega_{1}^{14\text{N}}/\omega_{1}^{1\text{H}}$ matching condition using Eq. \ref{Skrynikov}}. }
\begin{tabular}{lccc}
\hline
\textbf{Molecule-14N} & $\mathbf{\omega_{rf}^{14N}}$ \textbf{(ppm)} & $\mathbf{J_{HN}}$ \textbf{(Hz)} & $\mathbf{T_1^{14N}}$ \textbf{(ms)} \\
\hline
Uracil-N1/N3          & 130/110$^*$  & $68.4 \pm 2.1$/$69.2 \pm 1.9$   & $0.187 \pm 0.006$/$0.206 \pm 0.006$ \\
Thymine-N1/N3         & 120/100$^*$  & $76.0 \pm 5.0$/$61.9 \pm 1.8$   & $0.138 \pm 0.008$/$0.186 \pm 0.006$ \\
Adenine-N9/N10        & 120/30       & $81.9 \pm 1.0$/$72.1 \pm 1.0$   & $0.253 \pm 0.005$/$0.096 \pm 0.005$ \\
Cytosine-N1/N2        & 130/60       & $73.4 \pm 3.2$/$58.0 \pm 3.2$   & $0.262 \pm 0.015$/$0.143 \pm 0.015$ \\
Tryptophan-N1         & 120          & $53.0 \pm 19.0$                   & $0.073 \pm 0.020$                  \\
Urea-N1               & 36           & $66.0 \pm 3.3$                   & $0.116 \pm 0.0042$                  \\
Pyrrole-N1            & 106          & $34.7 \pm 0.22^{||}$                 & $0.935 \pm 0.017^{||}$                  \\
4-methylimidazole-N1/N2 & 167    & $5.19 \pm 0.18$/$5.94 \pm 0.18$ & $0.562 \pm 0.024$/$0.507 \pm 0.024$ \\
\hline
\multicolumn{4}{l}{\footnotesize\shortstack[l]{$^*$carrier set at the average value;$^{||}$ unrealistic fitted values as the measured values lead to $2\pi J_{HN} T_{1}^{14\text{N}}\approx 2$ (see text)}}
\end{tabular}
\label{table:fitted_data}
\end{table*}

\begin{figure}[t]
  \centering
  \includegraphics[page=5,trim=250pt 40pt 250pt 40pt, clip, width=0.5\textwidth]{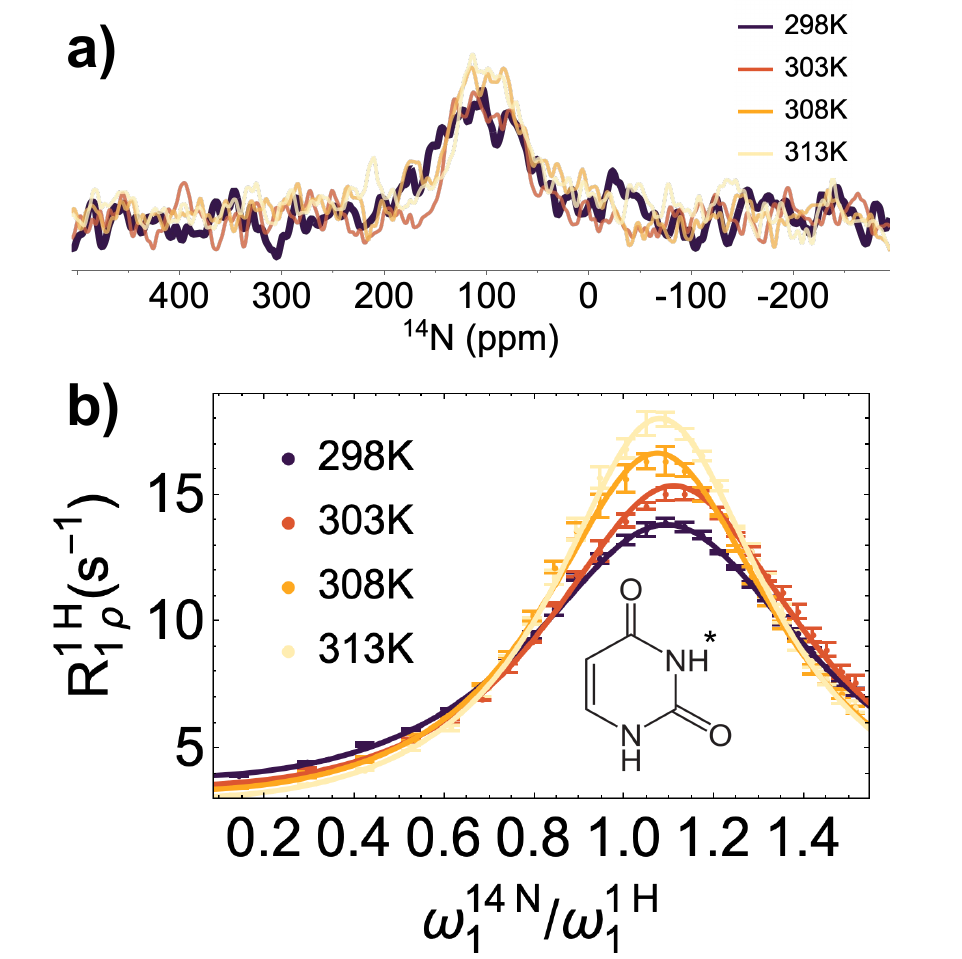}\caption{a) $^{14}$N spectra measured on uracil in DMSO-d6 using the aring sequence\cite{wang_triple-pulse_2021} averaged over 8192 transients at increasing temperatures from 298 K to 313 K (a line-broadening factor of 200 Hz was used). b) Rotating-frame $^1$H relaxation rate dispersion profiles as a function of the $\omega_{1}^{14\text{N}}/\omega_{1}^{1\text{H}}$ matching condition measured at increasing temperatures from 298 K to 313 K for the marked proton of uracil in DMSO-d6. Proton detection via the SR2K effect is significantly more sensitive to temperature changes than direct detection of $^{14}$N spectra. Points represent experimentally-derived $^1$H relaxation rates (error bars shown) and lines represent fits based on Eq. \ref{Skrynikov}. }
  \label{uracil_temp}
\end{figure}

Next, we explored the effect of temperature changes upon the SR2K contribution (Fig. \ref{uracil_temp}) and contrast it with the poor direct detection of $^{14}$N spectra for uracil in DMSO. As the temperature increases, the relaxation dispersion profile of $R_{1\rho}^{1\text{H}}$ increases in amplitude and becomes narrower around the Hartmann-Hahn condition consistent with a lengthening of the $^{14}$N polarization lifetime  as the  molecular tumbling increases. The extracted T$_{1}^{14\text{N}}$ increased from 0.224 $\pm$ 0.003 ms to 0.292 $\pm$ 0.004 ms, while the fitted scalar coupling varied within 2 Hz. Notably, the SR2K contribution to $^1$H relaxation grows as molecular tumbling accelerates, an inverted temperature dependence reminiscent of, but mechanistically distinct from, spin-rotation relaxation\cite{hubbard_theory_1963,spiess_rotation_1978}. 

\begin{figure}[t]
  \centering
  \includegraphics[page=4,trim=80pt 90pt 120pt 40pt, clip, width=0.5\textwidth]{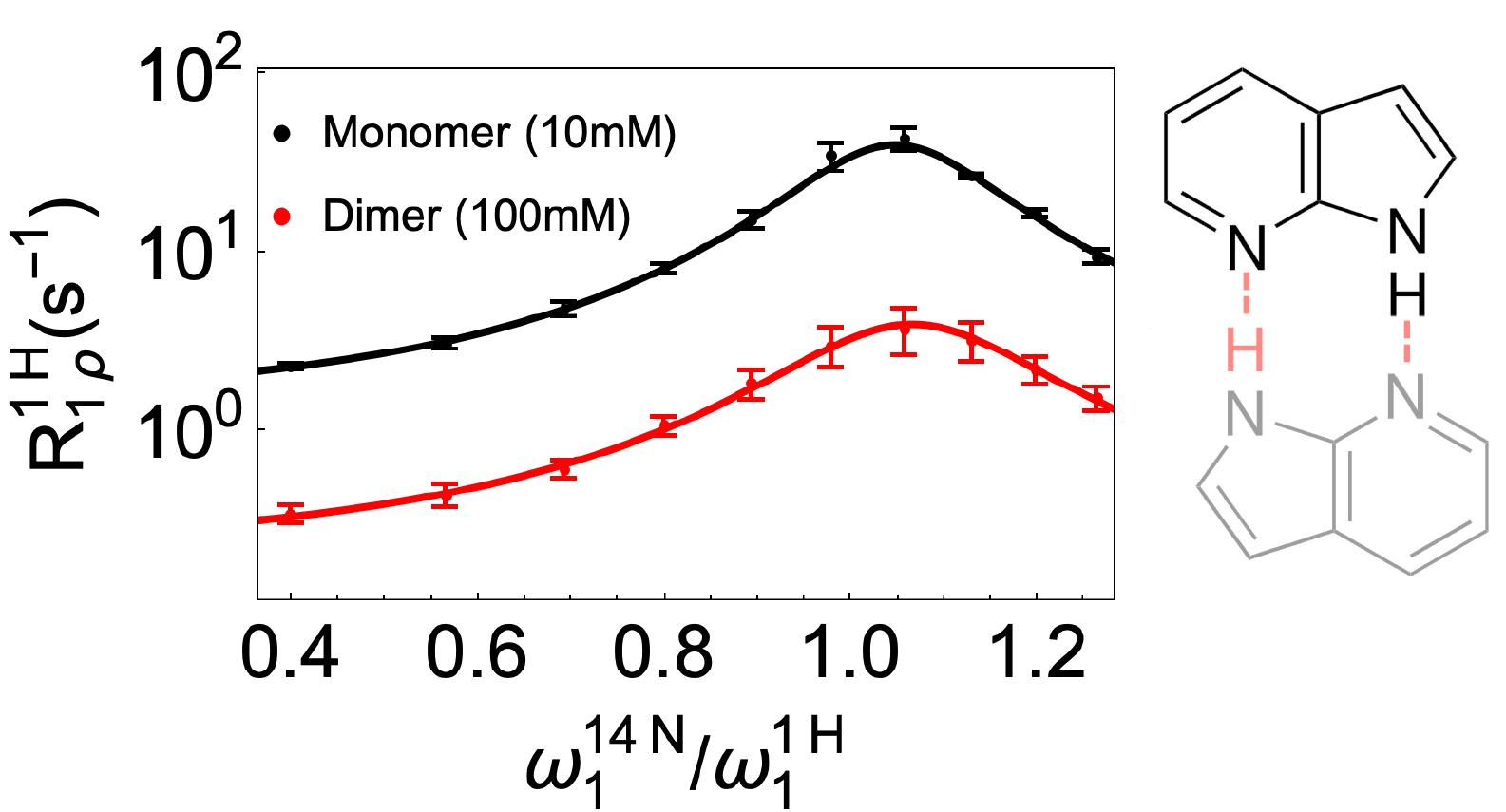}\caption{Rotating-frame $^1$H relaxation rate dispersion profiles of the pyrrolic proton (log-scale) as a function of the $\omega_{1}^{14\text{N}}/\omega_{1}^{1\text{H}}$ matching condition measured at increasing concentration of 7-azaindole in deuterated chloroform. Upon dimerization via hydrogen bonding, $^{14}$N polarization lifetime marginally decreases from $T_{\text{1}}^{\text{14N}}=0.7 \pm0.03$ ms at 10mM to $T_{\text{1}}^{\text{14N}}=0.58 \pm0.03$ ms at 100mM, while scalar coupling significantly reduces from $J_{HN}=64.9\pm0.8$ Hz to $J_{HN}=22.3\pm0.4$ Hz. Points represent experimentally-derived $^1$H relaxation rates (error bars shown) and lines represent fits based on Eq. \ref{Skrynikov}. }
  \label{azaindole}
\end{figure}

We further use the proton scalar relaxation enhancement to investigate intermolecular hydrogen bonding with N--H···N motifs, where the first nitrogen is covalently bonded, while the latter acts as an electron pair donor. The asymmetry of this three spin system suggests that, given sufficient frequency separation between the two nitrogen peaks, one can selectively reintroduce the SR2K effect and unequivocally probe the simultaneous coordination of the proton to both $^{14}$N nuclei, circumventing the need of isotopic replacement with $^{15}$N. However, the structural asymmetry also leads to very different scalar coupling constants. We investigated the well-studied\cite{rao_natural_1985} concentration-dependent 7-azaindole dimerization in deuterated chloroform in an attempt to probe proton bonding to both the pyrrolic and pyridinic nitrogen nuclei displaying different chemical shifts (Fig. \ref{azaindole}). Despite the large $^{14}$N chemical shift difference ($\Delta \delta \approx 130$ ppm) which makes this N--H···N system attractive for independent reintroduction of the two SR2K contribution by selective nitrogen irradiation, the scalar coupling constants differs by a factor of 35 (from DFT simulations), making the SR2K effect induced by the pyridinic nitrogen around 1000 times smaller than the pyrrolic one. Thus, we only investigated the proton scalar relaxation enhancement originating from the covalently bonded pyrrolic nitrogen as a function of concentration. The proton chemical shift changes from $\delta_H = 9.14$ ppm at 10 mM, which we attribute to the monomer form, to $\delta_H = 10.88$ ppm at 100 mM corresponding to dimer formation. Fitting the rotating-frame proton relaxation dispersion profile using Eq. \ref{Skrynikov} unveils that the $^{14}$N polarization lifetime decreased by about 20\%, while the one-bond scalar coupling constant dropped three-fold (Fig. \ref{azaindole}). This observation correlates with the expected retardation of molecular rotation modulating the nitrogen quadrupolar interaction, as well as with an elongation of the pyrrolic N-H bond due to intermolecular hydrogen bonding, that leads to a significant reduction of the scalar coupling.

\begin{figure}[h]
  \centering
  \includegraphics[page=6,trim=200pt 20pt 200pt 40pt, clip, width=0.5\textwidth]{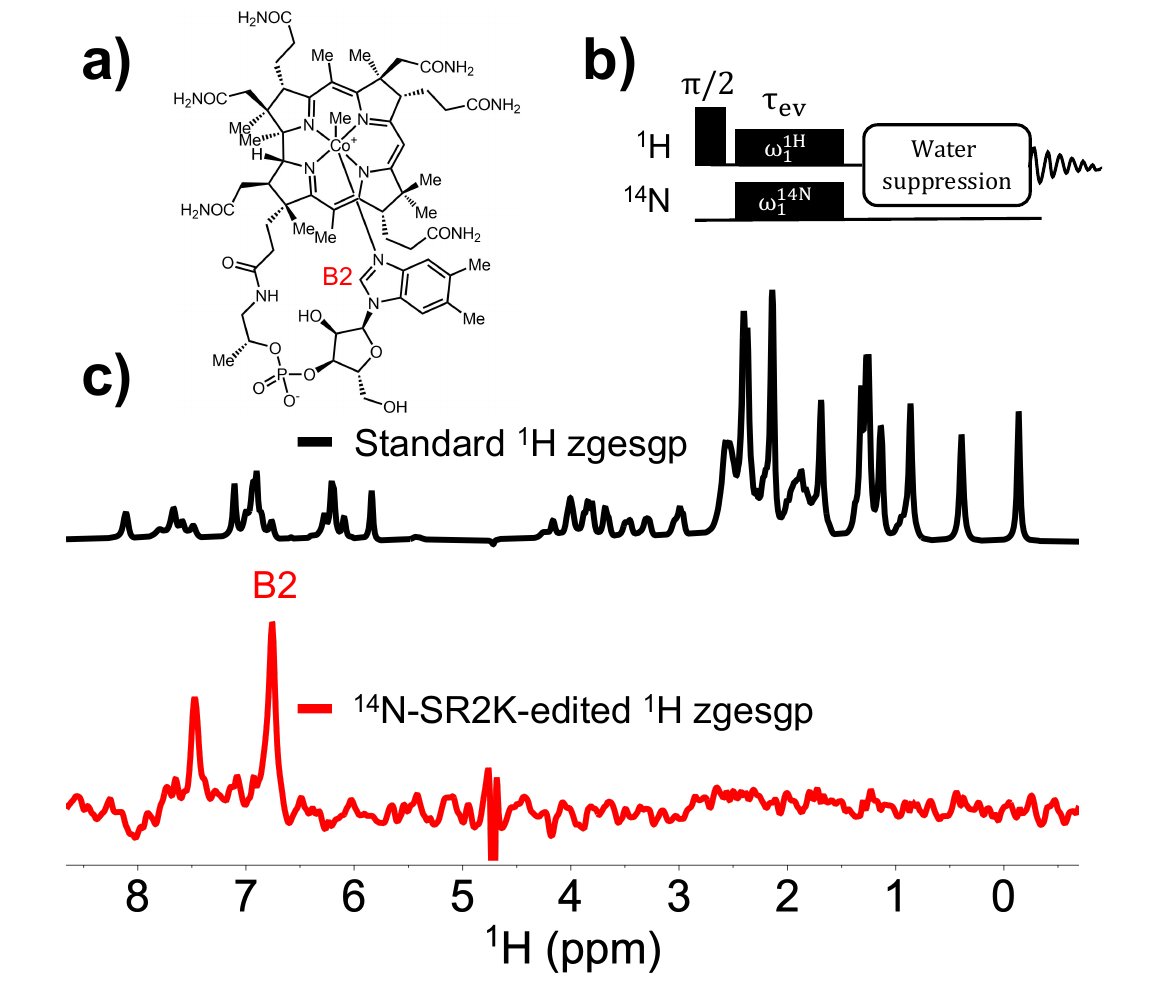}\caption{Relaxation-editing via $^{14}$N-SR2K effect in water-suppression proton experiments of a 5mM methylcobalamine in H$_2$O/D$_2$O=9/1 sample. a) Molecular structure of methylcobalamine with the imidazolic proton labeled B2 in red. b) General pulse sequence scheme showing how the double resonance spin-locking module inducing $^1$H-$^{14}$N SR2K can be incorporated into standard NMR methods such as water suppression. Here, we used the excitation sculpting (es) technique with gradient pulses (gp). c) Comparison between standard $^1$H spectrum acquired with water suppression sequence (black) and relaxation-edited $^1$H spectrum (red) derived as a difference between the on- and off-resonant $^{14}$N irradiation of imidazolic nitrogen atoms in methylcobalamine using the pulse sequence from b). The proton peak showing the highest SNR is the imidazolic proton B2 experiencing the cumulative scalar relaxation enhancement from the two neighboring nitrogen atoms.}
  \label{metcob}
\end{figure}

Finally, we demonstrate the utility of the $^1$H-$^{14}$N SR2K effect in relaxation-editing by incorporating the double resonance spin-lock module in a proton-detected water suppression experiment (Fig. \ref{metcob}). We opted for an aqueous solution of methylcobalamine with known assignment\cite{summers_complete_1986,gates_solvent_2024} focusing on the imidazolic proton labeled "B2" (Fig. \ref{metcob}a). We modified a standard water-suppression experiment, excitation sculpting with gradient pulses (zgesgp), by adding the dual-channel $^1$H-$^{14}$N spin-lock module right after the initial proton 90 degree pulse (Fig. \ref{metcob}b). Compared to the standard $^1$H spectrum with water suppression (Fig. \ref{metcob}c black), our approach allows filtering the proton nuclei J-coupled to fast-relaxing nitrogen atoms. Using the pulse sequence from Fig. \ref{metcob}b and taking the difference between the $^1$H-detected spectra with on- and off-resonant $^{14}$N spin-lock irradiation, the imidazolic proton peak B2 around 6.76 ppm can be separated out as its intensity  significantly reduced due to scalar relaxation enhancement during on-resonant irradiation of neighboring imidazolic nitrogen nuclei resonating at around 110 ppm. Another proton peak around 7.5 ppm was also filtered out and most likely corresponds to a slow-exchanging amide proton.

\section*{Conclusion}
\label{conclusion}
	
We show here that scalar relaxation of the second kind can be deliberately reintroduced into the rotating-frame relaxation of $^1$H nuclei through double-resonance spin-locking near the Hartmann-Hahn conditions for molecules of biological relevance. Scanning irradiation offsets and irradiation power ratios allows  indirectly characterizing $^{14}$N nuclei that are otherwise undetectable 
by direct NMR spectroscopy. The method simultaneously yields the $^{14}$N chemical shift, one- and two-bond $^1$H–$^{14}$N scalar coupling constants, and the longitudinal $^{14}$N polarization lifetime from a single set of proton rotating-frame relaxation dispersion measurements (within one transient for 20mM samples), exploiting the high sensitivity of $^1$H detection. 

Validation across ten molecular systems spanning nucleobases, amino acid side chains, and small nitrogen-containing heterocycles confirms the generality of the approach, with SR2K-driven $^1$H relaxation enhancements ranging from approximately 2-fold in amido protons of pyrimidines to over 20-fold in pyrrole. Remarkably, the detection of two-bond SR2K contributions in 4-methylimidazole shows that the method is not restricted to NH-bearing sites and extends to systems where direct proton–nitrogen bonding is absent, with the cumulative contribution from multiple $^{14}$N neighbors providing an additional fingerprint of the local nitrogen environment. These results establish SR2K-based rotating-frame relaxation as a simple, versatile and experimentally accessible tool for natural-abundance $^{14}$N characterization in relevant biomolecular systems that can be integrated naturally into existing multinuclear NMR workflows without isotopic enrichment with potentially even providing image contrast \textit{in-vivo}.

Lastly, we highlight that the approach to extracting spectral and dynamic parameters  through the SR2K effect used here is not limited to $^1$H-$^{14}$N pairs and could be reintroduced for any pair of scalar-coupled nuclei where one nucleus relaxes much faster than the observable one within the validity range of Eq. \ref{Skrynikov_cond}.

\section*{Experimental Details}
All nucleobases and nitrogen-containing small molecules used in this study were ordered from Sigma-Aldrich. The deuterated dimethyl sulfoxide (DMSO-d6, 99.8 atom\% D) was ordered from Thermo Fisher, while acetone-d6 (99.8 atom\% D) was ordered from VWR International. The four nucleobases (adenine, thymine, cytosine, and uracil) were dissolved in DMSO-d6 with a concentration of 20 mM. Guanine was not included in this study because of its low solubility in DMSO-d6. In addition, other small molecules containing NH or NH$_2$ functional groups were chosen based on previous $^{14}$N NMR studies\cite{deng_14n_2021}, such as 20mM pyrrole in acetone-d6, 40 mM 4-methylimidazole in acetone-d6 and 3 M urea in DMSO-d6. All nucleobases show very poor $^{14}$N spectra, while the last three samples display clear $^{14}$N signals (see Fig. 7). 

The experiments were performed on a Bruker AVANCE III 400 MHz NMR spectrometer with a 5 mm BBFO probe and a GREAT 3/10 XYZ gradient amplifier. The 1D $^1$H NMR spectra were acquired using the Bruker 1D zg pulse sequence to determine the chemical shift of the amino group. The size of FID is 16384 and the number of scans is 1. The 1D $^{14}$N NMR spectra were measured using aring pulse sequence\cite{wang_triple-pulse_2021} to reduce the acoustic ringing associated with low frequency measurements. The size of FID is 4096 and the number of scans is 8192. The pulse duration was optimized before the measurements of each sample. In the case of nucleobases which show very poor $^{14}$N peaks, the 3M urea in DMSO sample was used to calibrate the $^{14}$N pulse.

\begin{table*}[t]
\centering
\caption{Experimental parameters used for $^{1}$H-detected rotating-frame relaxation dispersion}
\begin{tabular}{ccccc}
\hline
\textbf{Molecule-\textsuperscript{1}H} & \textbf{$C_\text{M}$ (mM)} & \textbf{Solvent} & \textbf{$\omega_\text{rf}^\text{1H}$ (ppm)} & \textbf{$\omega_1^\text{1H}$ (kHz)} \\
\hline
Uracil-H1/H3          & 20   & DMSO-$d_6$    & 11.01/10.81$^*$ & 2   \\
Thymine-H1/H3         & 20   & DMSO-$d_6$    & 10.99/10.58$^*$ & 2   \\
Adenine-H9/H10        & 20   & DMSO-$d_6$    & 12.85/7.09      & 2/3 \\
Cytosine-H1/H2        & 20   & DMSO-$d_6$    & 10.32/6.95      & 3   \\
Tryptophan-H1         & 20   & DMSO-$d_6$    & 10.90           & 3   \\
Urea-H1               & 3000 & DMSO-$d_6$    & 5.63            & 2   \\
Pyrrole-H1            & 40   & Acetone-$d_6$ & 6.79            & 1   \\
4-methylimidazole-H1/H2 & 40 & Acetone-$d_6$ & 7.48/6.73       & 2   \\
azaindole-H1 & 10/100 & CDCl$_3$ & 9.14/10.88 & 2 \\
methylcobalamine-B2 & 5 & H$_2$O/D$_2$O = 9/1& 6.76 & 2 \\
\hline
\multicolumn{4}{l}{$^*$ Carrier set at the average of these values.}

\end{tabular}
\label{tab:nmr_params}
\end{table*}

\begin{figure}[t]
  \centering
  \includegraphics[page=7,trim=30pt 00pt 280pt 00pt, clip, width=0.5\textwidth]{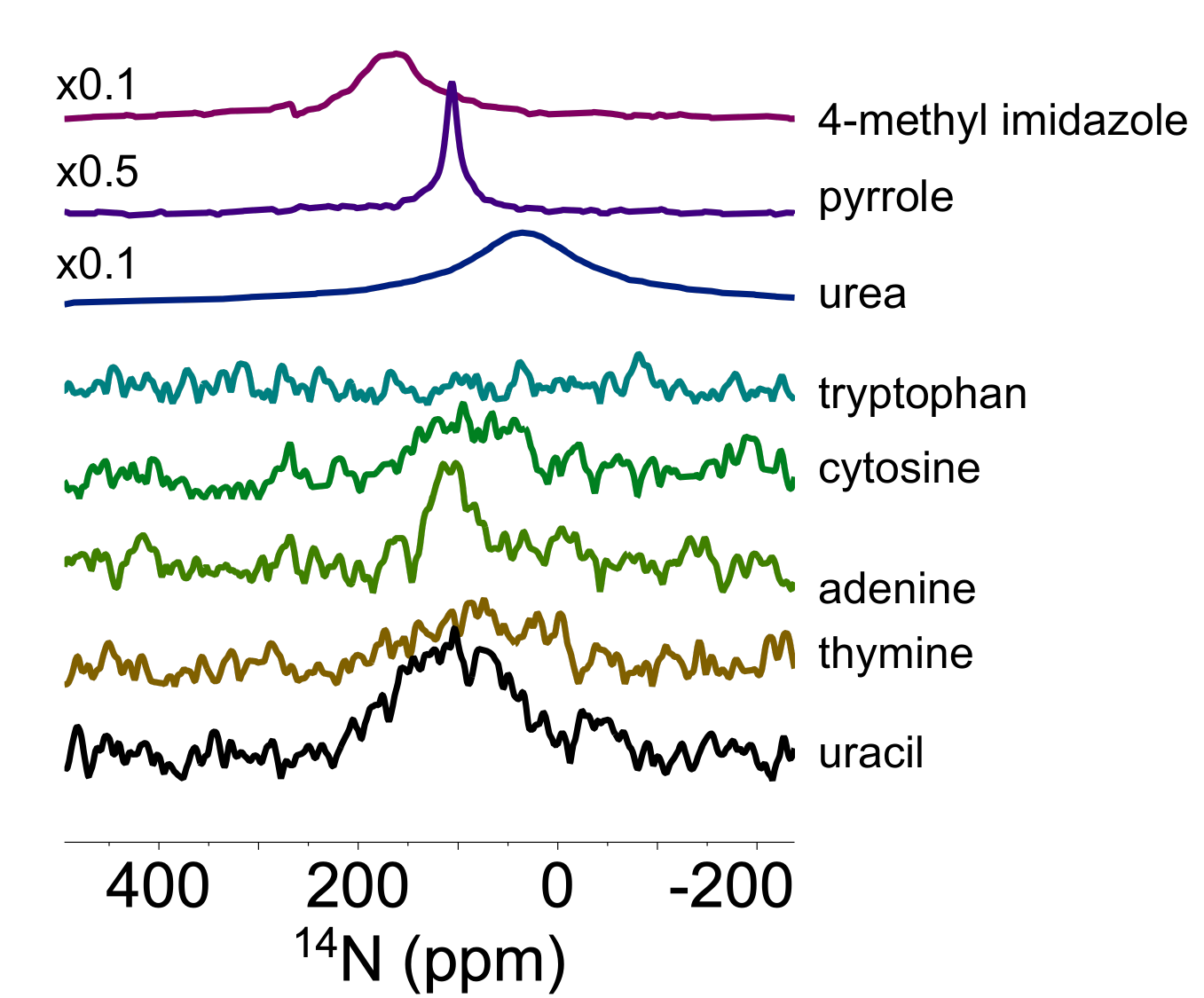}\caption{$^{14}$N NMR spectra measured using aring pulse sequence\cite{wang_triple-pulse_2021} with 8192 transients for the chemical systems described in Table II.}
  \label{azaindole}
\end{figure}

The longitudinal relaxation rate (R$_1$) was measured using the inversion recovery pulse sequence (t1ir) in order to properly set the recovery delay. The $^1$H relaxation rate in the rotating frame (R$_{1\rho}$) experiments were performed using a modified spin lock pulse sequence (see Fig. 2C inset). After the initial recovery delay at the beginning of each transient, a 90° excitation pulse was applied on the Y axis of $^1$H channel only, followed by a double-resonance spin-locking irradiation along X axis. The continuous wave irradiation was applied simultaneously on both $^1$H channel and the $^{14}$N channel to lock transverse magnetization, followed by the acquisition period. The transmitter offset frequency of the $^1$H channel was set at the resonance position of the targeted NH or NH$_2$ proton peaks, while the $^{14}$N carrier was scanned by monitoring the largest drop of the proton signal for a constant irradiation time of 0.1 s. The irradiation power on the $^1$H channel for each experiment is given in Table II. In order to extract $^1$H-$^{14}$N scalar couplings and $^{14}$N polarization lifetimes, multiple experiments with increasing irradiation power on the $^{14}$N channel were performed (see Fig. 3), from which the $^1$H relaxation rates in the rotating frame (R$_{1\rho}^{1\text{H}}$) were extracted and then fitted with Eq. 1. The free fitting variables were the proton rotating-frame relaxation rate under no $^{14}$N irradiation power ($R_{1\rho}^{0}$), the $^1$H-$^{14}$N scalar coupling constants ($J_{HN}$) and the $^{14}$N longitudinal polarization lifetime ($T_{1}^{14N}$). In some cases, we've added an extra fitting variable to account for the slight deviation from the Hartmann-Hahn condition of the maximum SR2K contribution. We attribute these deviations to either pulse imperfections or chemical exchange phenomena. We performed double resonance spin locking experiments on uracil at 4 different temperatures and compare the temperatures changes in $^{14}$N spectra with the ones in R$_{1\rho}^{1\text{H}}$ profiles (see Fig. 4).

Concentration-dependent dimerization of 7-azaindole experiments have been carried in deuterated chloroform at two concentrations of 10mM and 100mM (Fig. 5). Two nitrogen peaks have been observed in $^{14}$N spectra at 90 ppm and 230 ppm corresponding to the pyrrolic and pyrimidinic nitrogen nuclei, respectively. The nitrogen spin-lock carrier was set on the pyrrolic peak only, as irradiating the pyrimidinic nitrogen did not lead to any significant SR2K contribution for the $^1$H involved in the intermolecular hydrogen bond.

For experiments on methylcobalamine, we've prepared a 5mM sample in H$_2$O/D$_2$O=9/1 solvent and track the SR2K effect induced by the imidazolic nitrogen nuclei to the neighboring B2 proton (Fig. 6). We have used the excitation sculpting with gradient pulses (zgesgp) for water suppression to which we added the $^1$H-$^{14}$N double-resonant spin-lock module (Fig. 6b) with $^1$H carrier set at 6.76 ppm based on previous assignment, while the optimal $^{14}$N carrier was find at 110 ppm by optimization as described in Fig. 2. The relaxation-editing strategy to isolate the B2 signal is based on the different decay rates of protons coupled to fast-relaxing nitrogen nuclei during the spin-lock period $\tau_{ev}=0.2$ s with on- or off-resonant $^{14}$N irradiation. The $^{14}$N-SR2K-edited spectra is computed as the difference between the on- and off-resonant $^{14}$N irradiation experiments. Four transients have been acquired for the standard water suppression experiment and 16 transients for the $^{14}$N-SR2K-edited variant.

\section*{Acknowledgements}

F.T. acknowledges funding from National Medical Project MySMIS 326475 and Project ELI-RO/RDI/2024/14 SPARC funded by the Institute of Atomic Physics (Romania). A.J. acknowledges funding from the US National Science Foundation, award no. CHE 2505792 and an award from the ACS PRF 68117-ND6, as well as an unrestricted gift from Google.

\section*{Supporting information}
Supporting Information: Experimental protocols, $^{14}$N NMR spectra for all molecular systems presented in Fig. \ref{all_molec} using aring sequence\cite{wang_triple-pulse_2021}; Table summarizing experimental parameters used for the double resonance spin-lock experiments; relaxation dispersion data and Mathematica scripts used for fitting are provided as well.

\bibliography{references.bib}

\end{document}